\documentclass[aps,prb,twocolumn,amsmath,amssymb,nofootinbib,superscriptaddress]{revtex4-2}

\usepackage{amsmath}
\usepackage{amssymb}
\usepackage{graphicx}
\usepackage{array}
\usepackage{bm}
\usepackage{braket}
\usepackage{booktabs}
\usepackage{dsfont}
\usepackage{dcolumn}
\usepackage{esint}
\usepackage{epstopdf}
\usepackage{mathrsfs}
\usepackage{enumitem}

\usepackage[colorlinks=true,citecolor=green,linkcolor=blue]{hyperref}
\usepackage{color}

\begin{document}

\title{Scheme for braiding Majorana zero modes in vortices using STT-matrix}

\author{G. -Y.~Huang}
\affiliation{College of Computer Science and Technology, National University of Defense Technology, Changsha 410073, China}
\author{J.-B.~Fu}
\affiliation{College of Computer Science and Technology, National University of Defense Technology, Changsha 410073, China}
\author{X.-F.~Yi}
\affiliation{College of Computer Science and Technology, National University of Defense Technology, Changsha 410073, China}
\author{W.-C.~Wang}
\affiliation{College of Computer Science and Technology, National University of Defense Technology, Changsha 410073, China}
\author{B.~Ren}
\affiliation{College of Computer Science and Technology, National University of Defense Technology, Changsha 410073, China}
\author{Z.-H.~Yang}
\affiliation{College of Computer Science and Technology, National University of Defense Technology, Changsha 410073, China}
\author{S.-C.~Xue}
\affiliation{College of Computer Science and Technology, National University of Defense Technology, Changsha 410073, China}
\author{X.-F.~Zhang}
\email[Corresponding author: ]{xfz-6446@nudt.edu.cn}
\affiliation{College of Computer Science and Technology, National University of Defense Technology, Changsha 410073, China}
\author{M.-T.~Deng}
\email[Corresponding author: ]{mtdeng@nudt.edu.cn}
\affiliation{College of Computer Science and Technology, National University of Defense Technology, Changsha 410073, China}
\affiliation{Hefei National Laboratory, Hefei 230088, China}

\begin{abstract}
Majorana zero modes (MZMs), promising for topological quantum computation, are naturally hosted in vortices of two-dimensional topological superconductors (TSCs). However, precise control and braiding of these vortex-bound MZMs remains a significant challenge. This work proposes and numerically demonstrates a novel braiding scheme utilizing a programmable matrix of spin-transfer torque (STT) devices (STT-matrix) integrated with a TSC layer. By selectively activating individual STT elements, their localized stray fields enable deterministic manipulation, including driving, braiding, and fusion, of superconducting vortices and their associated MZMs. We establish a comprehensive simulation framework combining finite element analysis for STT-induced vortex formation, time-dependent Ginzburg-Landau equations for vortex dynamics and time-dependent Bogoliubov-de Gennes equations for MZM evolution. Simulations confirm the STT-matrix's capability for high-fidelity vortex manipulation and demonstrate MZM braiding dynamics. We quantify the impact of vortex acceleration and finite MZM coupling on braiding fidelity, showing it can be optimized by adjusting STT spacing and vortex separation. Furthermore, we demonstrate controlled MZM fusion and measure the resultant energy splitting. This STT-matrix-based approach offers a highly versatile, scalable, and potentially practical platform for operating MZMs within TSC vortices, advancing towards fault-tolerant topological quantum computation.
\end{abstract}

\date{\today}

\maketitle

\section{Introduction}\label{introduction}

The potential of quantum computation has sparked significant research interest in recent decades. However, achieving practical quantum computation necessitates the implementation of fault-tolerant techniques. A promising approach is topological quantum computation (TQC) ~\cite{ap303.2,rmp80.1083,bookIntroToTQC}. The cornerstone of TQC lies in topological excitations known as non-Abelian anyons, which emerge in many-body systems. These non-Abelian anyons form qubits that are topologically protected, rendering them resilient against local noise and enabling extremely low error rates that meet the threshold requirements for fault-tolerant quantum computation. In TQC, quantum gates are implemented through the braiding of non-Abelian anyons, referred to as braiding gates. Braiding results are obtained by measuring the signals associated with information after the fusion of non-Abelian anyons—a process involving coupling and hybridization of these anyons. This approach offers a promising route towards realizing fault-tolerant quantum computation and paves the way for more robust and reliable applications in quantum computing.

In recent years, the Majorana zero mode (MZM) has emerged as a prominent non-Abelian anyon for realization in solid-state systems~\cite{pu44.131, nphys5.614, arcmp4.113,rmp87.137, npjqi1.15001, nrm3.52, scpma64.107001}. Signatures of MZMs have been observed in one-dimensional (1D) systems, such as nanowires and ferromagnetic atomic chains~\cite{science336.1003, nl12.6414, nphys8.887, nature531.206, science354.1557, Vaitiekenas2020}. However, since MZMs cannot be braided within a single 1D structure, the creation of a 1D network becomes crucial for implementing braiding operations~\cite{nl19.218,prm2.093401}. In topological semiconductor-superconductor heterostructure networks, braiding can be achieved by locally controlling electron density using electrode gates~\cite{Alicea2011, Aasen2016, Karzig2017, Yuan2021}.

Compared to 1D or quasi-1D systems, two-dimensional (2D) topological superconductors (TSCs) offer a more natural platform for the braiding of anyons. Various proposals have been put forward to realize MZMs in 2D systems, including fractional quantum Hall effects~\cite{npb360.362}, intrinsic $p$-wave superconductivity~\cite{prl86.268}, topological insulators proximitized by s-wave superconductors~\cite{prl100.096407}, and 2D semiconductor/superconductor hybrid structures in proximity to magnetic insulators or subjected to external magnetic fields~\cite{prl104.040502, prb81.125318}. In particular, superconductor vortex is a natural host for MZM and other sub-gap states~\cite{Vaitiekenas2020, Deng2025}. MZM signatures within vortices have been observed in hybrid 2D systems~\cite{prl114.017001, prl116.257003} and iron-based superconductors over the past decade~\cite{Wang2018, prx8.041056, Kong2019, Zhu2019, Zhu2021, Kong2021nc, Li2022, Ge2023}, opening up new avenues for topological quantum computation with vortices.

The observation of 2D MZMs has stimulated extensive investigations into corresponding studies on braiding methods~\cite{prb79.205102, prb96.184508, prb101.024514, prr2.023205, jpd54.424003, prb105.014507, prb103.054504, prb96.035444, prl122.146803, prl128.016402}. For example, various approaches have been proposed including nanoscale tips or tip-shaped pinning sites~\cite{prb96.184508, prb101.024514}, electrically/magnetically controlled braiding techniques~\cite{prr2.023205, jpd54.424003, prb105.014507, arxiv2210.10650}, and optically mediated methods~\cite{ncomms7.12801, prb104.104501}. However, the majority of these methods fail to satisfy the essential criteria required for successful implementation of MZM braiding protocols. In particular, gate-controlled techniques relying on electrostatics are incompatible with 2D semiconductor-superconductor hybrids or iron-based superconductors due to the screening effect of metal electrodes, which prevents effective electron density tuning. Meanwhile, tip-based approaches face integration challenges.

\begin{figure}
\centering 
\includegraphics[width=8 cm]{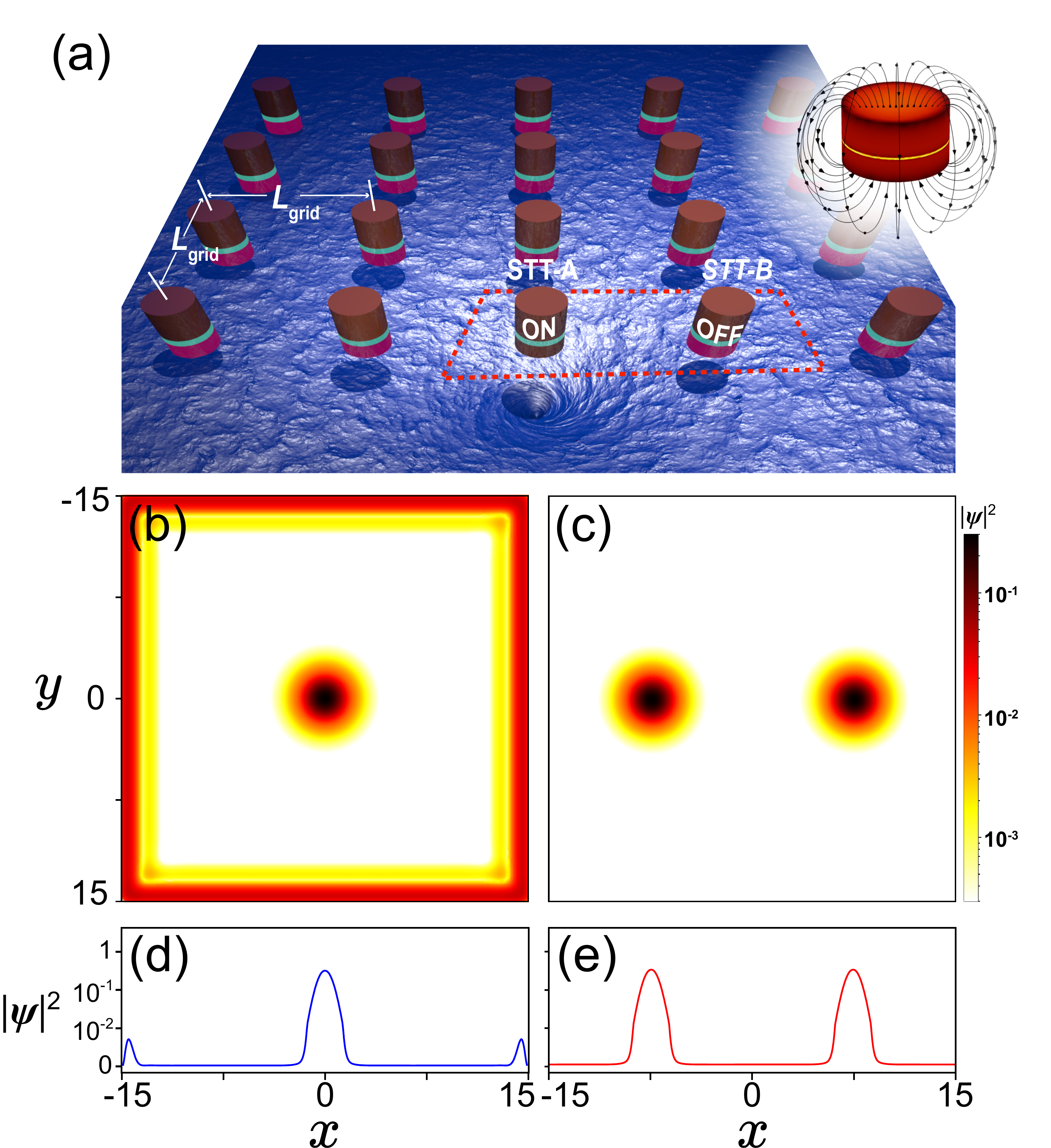}
\caption{\label{figDevice} 
(Color online) Device structures and MZMs in TSC. 
(a) Schematic diagram illustrating the device with two layers: the control layer, consisting of an STT-matrix, and the material layer, comprising the TSC. The STTs can be individually switched ON or OFF, as indicated by the red dashed rectangle. In the ON state, a vortex in the TSC layer can be driven or dragged via the local magnetic stray field generated by each active STT element. The inset shows the simulated stray field of an isolated STT. (b)-(c) Probability density distribution of MZMs in a TSC with one and two vortices, respectively. To highlight the MZM edge mode, logarithmic scale is used to display probability. Here, the TSC is modeled as a square geometry with open boundary conditions. (d)-(e) Line-cuts along the x-axis taken at $y=0$ corresponding to (b) and (c), respectively. Parameters used for calculations include $\tilde{W}=0.3$, $\tilde{v_F} = 0.5$, $\Delta_0 = 1$, $\xi = 0.25$, $d=1$, lattice size $L_x =L_y =30$.
} 
\end{figure}

In this study, we propose a novel braiding scheme to address the challenges of 2D MZMs using spintronics devices. Similar to the control elements in Field-Programmable Topological Array (FPTA)~\cite{arxiv2010.02130}, magnets with nanometer scales can be individually configured and controlled by electrical current pulses. We implement spin-transfer-torque (STT) devices, which have been extensively studied in the field of spintronics~\cite{rmp76.323}, to enable independent, simultaneous, and localized programmability of various properties including electromagnetic fields, synthetic spin-orbit fields, and superconductivity.

Leveraging STT devices for braiding offers significant advantages in terms of efficiency, control speed, and applicability compared to existing proposals. Previous studies by Fatin {\it et al.} in Ref.~\cite{prl117.077002} and Zhou {\it et al.} in Ref.~\cite{Zhou2019} have demonstrated the theoretical feasibility of manipulating and braiding MZMs using magnetic tunneling junctions (MTJs) in 2DEG-superconductor hybrids. Inspired by these groundbreaking works, we propose the utilization of STT-matrix as a novel approach to braid MZMs within vortex-type TSCs. It is noteworthy that our work differs from Ref.~\cite{prl117.077002} in addressing different material systems, employing different numerical methods, expanding to time-dependent simulations and braiding dynamics.

The article is structured as follows: Section~\ref{secBasis} provides an overview of the fundamental concepts and system structures of MZM braiding devices. In Section~\ref{secModel}, we present a detailed description of the model and method based on time-dependent Ginzburg-Landau (TDGL) theory and time-dependent Bogoliubov-de Gennes (TDBdG) equations, with particular emphasis on integrating the STT-matrix into the TDGL equations. In Section~\ref{secBraiding}, we simulate and demonstrate the manipulation of MZMs by the STT-matrix, including their ability to drive single vortices (Subsec.~\ref{subsecSingle}) and analyze braiding dynamics (Subsec.~\ref{subsecBraiding}). Additionally, we propose a fusion method for MZMs in the STT setup (\ref{subsecFusion}). Finally, conclusions are drawn in Section~\ref{secConclusion}. 


\section{Fundamental concept and architectural framework}\label{secBasis}

Our proposals are formulated based on three fundamental principles. Firstly, within TSCs, a vortex hosts an MZM. As mentioned in the first section, $p$-wave superconductor vortices have been proposed as one of the most promising platforms for hosting MZMs, and experimental evidence of MZMs in vortices has been reported. Secondly, a vortex can be individually manipulated by a local magnetic field. Previous studies have demonstrated the creation and manipulation of vortices using field coils on superconducting quantum interference devices~\cite{apl80.1010, apl93.172514, nl23.4669}. Thirdly, the local magnetic field can be provided by the stray field of STT device matrix~\cite{prl117.077002, Zhou2019}. Combining these principles implies that braiding the MZMs is equivalent to braiding the vortices in $p$-wave superconductors, which is further equivalent to configuring the STT-matrix.

An STT device typically comprises an MTJ multiple layer structure, consisting of two ferromagnetic layers separated by a thin non-magnetic spacer layer. One of the ferromagnetic layers, referred to as the ``reference layer," maintains a fixed magnetization direction during operation. The other ferromagnetic layer, known as the ``free layer", can have its magnetization direction altered by the spin-polarized current flowing through the structure. The change in magnetization direction of the free layer corresponds to the alteration of the STT state, which is achieved through the interaction between the angular momentum of polarized current and magnetic moments of the free layer~\cite{Ralph2008}. Depending on the direction of current, either alignment or anti-alignment can be established between the magnetization directions of the free layer and reference layer. Consequently, programmability of STT states arises from distinct net stray fields exhibited during alignment or anti-alignment. The size of the STT device can now be scaled down to a few tens of nanometers, enabling the integration of billions of these devices into a single chip. As a result, the STT-matrix can provide exceptional precision and scalability for vortex manipulation.

The STT-matrix, as shown in Fig.~\ref{figDevice}(a), functions as the control layer within the system. Each individual STT can be independently controlled using conventional addressing techniques~\cite{iedm1609379}. By altering the current direction, specific STTs can be selectively activated (ON) or deactivated (OFF), enabling precise alignment or anti-alignment of the free layer with the reference layer for manipulation and propulsion of vortices within the TSC layer. In Figs~\ref{figDevice}(b)-(c), we demonstrate MZMs within TSC vortices controlled by STTs (see formula in Sec.~\ref{subsecTSC}). When only one STT is activated, a vortex is induced by this particular STT, resulting in the emergence of a point-like MZM within the vortex. Since MZMs always appear in pairs, another MZM manifests as a 1D edge mode [Fig.~\ref{figDevice} (b) and (d)], akin to edge MZMs observed near a s-wave superconductor and ferromagnetic insulator in semiconductor systems~\cite{epl99.50004}. Conversely, activating two STT devices generates a pair of vortices hosting point-like MZMs [Fig.~\ref{figDevice} (c) and (e)].


\section{Model and Method}\label{secModel}

In this section, our objective is to establish the comprehensive framework model as described in the preceding section and devise a methodology for simulating MZMs braiding. 

Firstly, vortices are induced by STT-matrix, which is modeled via a Finite Element Analysis method (see Appendix~\ref{secAppendixFEA} and Ref.~\cite{Deng2024}). The vortex dynamics are governed by the TDGL equations, which are a powerful tool for studying the dynamics of vortex systems in type-II superconductors with mixed phases~\cite{bookIntroToSC}. 
While the TDGL theory was originally formulated for dirty or gapless superconductors near the critical temperature $T_c$~\cite{Schmid1966}, its generic phenomenological features enable its validity across a broader range of conditions. Specifically, for iron-based superconductors, where the penetration length $\lambda$ is typically several orders of magnitude larger than the coherence length $\xi$~\cite{PhysRevB.90.144517, PhysRevB.81.224520, PhysRevB.78.220510}, the dirty limit construction of TDGL is well-justified.
To incorporate the effect of the STT-matrix, we modify the TDGL equations by including local magnetic fields associated with it. This modification enables us to simulate and analyze individual vortex dynamics effectively~\cite{bookShortcut,cpc291.108799}.

The model of TSCs is an effective 2D representation. The MZMs are obtained by solving the static model of the TSCs. By incorporating vortex dynamics derived from TDGL equations, the braiding dynamics of MZMs can be solved using TDBdG equations. 

It should be noted that although we utilize a 2D model as an illustrative example in our simulations, we expect that our approach can also effectively capture the behavior of quasi-2D TSCs or the surface states of bulk topological materials.


\subsection{TDGL equations}\label{subsecTDGL}

The TDGL equations describe a system of nonlinear equations governing the behavior of the superconductor order parameter $\Psi$ and electromagnetic potential $(\Phi, \mathbf{A})$. These equations can be expressed as~\cite{aam115.63, bookShortcut}:

\begin{eqnarray}
  \label{eqTDGLDimension}
\frac{\hbar^2}{2mD}(\frac{\partial}{\partial t}+i\frac{q}{\hbar}\Phi)\Psi
& = & -\frac{1}{2m}(\frac{\hbar}{i}\nabla-q\mathbf{A})^2\Psi+\alpha\Psi\nonumber\\
&   & -\beta|\Psi|^2\Psi,\\
\sigma(\frac{\partial \mathbf{A}}{\partial t}+\nabla\Phi)
& = & \frac{q\hbar}{2mi}(\Psi^*\nabla\Psi-\Psi\nabla\Psi^*)\\
&   & -\frac{q^2}{m}|\Psi|^2 \mathbf{A}-\frac{1}{\mu_0}\nabla\times(\nabla\times \mathbf{A}-\mathbf{B}).\nonumber
\end{eqnarray}
where $\hbar$ represents the reduced Planck's constant, $m$ denotes the mass of a Cooper pair, and $q=2e$ signifies the charge carried by a Cooper pair (where $e$ is the elementary charge carried by an electron). The parameters $\alpha$ and $\beta$ correspond to the phenomenological aspects of Gibbs free energy in the GL theory. Specifically, $\alpha(T)$ is a temperature-dependent parameter given by $\alpha(0)(1-T/T_c)$, where $T_c$ represents the critical temperature of the superconductor. On the other hand, $\beta$ remains as a constant value. Additionally, $D$, $\sigma$, and $\mu_0$ are material-specific parameters: $D$ refers to diffusion coefficient; $\sigma$ corresponds to conductivity; and finally, $\mu_0$ stands for permeability of free space. Furthermore, we consider an external magnetic field denoted as 	$\mathbf{B}=\mathbf{B}(x,y,t)$ which varies with time and can be controlled using the STT-matrix.

Related to GL equations, there are two important characteristic lengths: the London penetration depth, denoted as $\lambda=\sqrt{{m\beta}/(q^2 \mu_0 \alpha)}$, and the GL coherence length, denoted as $\xi={\hbar}/{\sqrt{2m\alpha}}$. The ratio of these two lengths, referred to as the Ginzburg-Landau parameter, $\kappa={\lambda}/{\xi}$, serves as an indicator for distinguishing between type I superconductors ($\kappa<1/\sqrt{2}$) and type II superconductors ($\kappa>1/\sqrt{2}$).

The dimensionless coordinates and time are rescaled by $\lambda$ and $t_0=\xi^2/D$, respectively. 
Consequently, the dimensionless derivatives can be expressed as 
\begin{eqnarray}
\tilde{\nabla}=\lambda \nabla,\quad 
\partial_\tau=t_0 \partial_t.
\end{eqnarray}
The dimensionless quantities can be defined by introducing the corresponding unit,
\begin{eqnarray}
\begin{aligned}
\Psi_0 &= \sqrt{\frac{\alpha}{\beta}}, &
\sigma_0 &= \frac{1}{\mu_0 D \kappa^2}, &
\Phi_0 &= \frac{\hbar D \kappa}{q\xi^2},  \\
\mathbf{A_0} &= \frac{\hbar}{q\xi}, &
\mathbf{B_{0}} &= \frac{\hbar}{q \kappa \xi^2}, &
\end{aligned}
\end{eqnarray}

and we have
\begin{eqnarray} \begin{aligned}
  \psi &=\Psi/\Psi_0,~                             &
  \tilde{\sigma} &=\sigma/\sigma_0,              &
  \phi &=\Phi/\Phi_0,\\
  \tilde{\mathbf{A}} &=\mathbf{A}/\mathbf{A_0},   &
  \mathbf{b} &=\mathbf{B}/\mathbf{B_{0}}. &
\end{aligned} \end{eqnarray}

After performing algebraic calculations and applying a gauge transformation (see~\ref{secAppendixA} for detailed procedures), we can obtain a dimensionless form as follows:
\begin{eqnarray}
\partial_\tau \psi & = & -(\frac{i}{\kappa}\nabla+\mathbf{A})^2\psi+\psi-|\psi|^2\psi,\\
\sigma \partial_\tau \mathbf{A} & = & \frac{1}{2i\kappa}(\psi^*\nabla\psi-\psi\nabla\psi^*)-|\psi|^2 \mathbf{A}\nonumber\\
& & -\nabla\times(\nabla\times \mathbf{A}-\mathbf{b}),
\end{eqnarray}
where the redefined variables $\tilde{\nabla}$, $\tilde{\sigma}$ and $\tilde{\mathbf{A}}$ are now denoted as $\nabla$, $\sigma$ and $\mathbf{A}$.

Extended boundary conditions are employed, specifically designed for numerical simulations~\cite{bookShortcut}, which involve enlarging the superconductor with a non-superconducting and insulating material. Mathematically, the TDGL equations are modified by introducing a region-dependent function. Consequently, the dimensionless boundary conditions can be expressed as
\begin{eqnarray}
\psi|_{\partial \Omega} & = & 0,\\
\nabla\times \mathbf{A}|_{\partial \Omega} & = & \mathbf{b}|_{\partial \Omega},
\end{eqnarray}
in which $\partial \Omega$ represents the external boundary, and $\mathbf{b}$ is the dimensionless magnetic field. 
The initial condition for the TDGL equations is
\begin{eqnarray}
\psi|_{\tau=0}=1,\mathrm{\, \in\, superconductor\, region }.
\end{eqnarray}

Given the boundary and initial conditions, the quantities $\psi=\psi(t)$ and $\mathbf{A}=\mathbf{A}(t)$ can be determined by solving the TDGL equations. Here, our primary focus lies on the observable quantity, namely the Cooper pair density $\rho$, which can be expressed as $\rho=|\Psi|^2=|\psi|^2\Psi_0^2$.

\subsection{Model of the STT-matrix}

In the previous subsection, we present the formalism of TDGL theory. By imposing appropriate boundary and initial conditions, numerical methods can be employed to obtain the dynamics of vortices. Unlike prior studies on vortex lattices formed in uniform external magnetic fields, our focus here lies on the vortex lattice generated by the STT-matrix, in order to achieve individual vortex manipulation. To model the STT-matrix, we abstract it as a series of localized magnetic fields acting upon the superconductor. Through integration with the STT-matrix, the TDGL equations enable simulation of vortex dynamics driven by STTs.

The localized magnetic field generated by an activated STT device can be approximately represented by a Gaussian magnetic field, which corresponds to the external magnetic field $b(x,y)$ in the TDGL equations,
\begin{eqnarray}
\label{eqbSTT}
b_{i,j}(x,y)=b_{0}e^{-[(x-X_{i,j})^2+(y-Y_{i,j})^2]/R_v^2},
\end{eqnarray}
where $b_{0}$ is the maximum magnetic field, $(X_{i,j}, Y_{i,j})$ is the central coordinate of the STT in the $i^{\rm th}$ row and the $j^{\rm th}$ column. In the case of a Gaussian-shaped magnetic field, $R_v$ represents the characteristic radius of an individual STT's magnetic field. It should be noted that while selecting a Gaussian-shaped magnetic field for each STT is not crucial, any model that can capture the localized nature of the STT's magnetic field can be employed for vortex control.

The distance between the nearest STTs in the STT-matrix, denoted as $L_{grid}$, is defined as $L_{grid} = X_{i,j}-X_{i-1,j}=Y_{i,j}-Y_{i,j-1}$ for a square STT-matrix, as illustrated in Fig.~\ref{figDevice}(a). The net magnetic field from the entire STT-matrix, denoted as $b_{\rm net}$ at time $\tau$, is the aggregate of all individual STT magnetic fields,
\begin{eqnarray}
b_{\rm net}(\tau)=\sum_{i,j} s_{i,j} (\tau) b_{i,j}(x,y).
\end{eqnarray}
The time dependence arises from the status function $s_{i,j} (\tau)$, which is a binary function representing the ON or OFF state of the $i,j$-\textit{th} STT,
\begin{eqnarray}
s_{i,j} (\tau) = 
\begin{cases}
1, & \mathrm{ON}\\
0, & \mathrm{OFF}
\end{cases}.
\end{eqnarray}
The status function $s_{i,j} (\tau)$ serves as the crucial link between the control layer and the magnetic field texture, effectively demonstrating the programmable nature of the STT-matrix.

\subsection{Model of TSCs and the TDBdG equations}
\label{subsecTSC}

To describe the TSCs, we employ the 2D effective model, which offers the advantage of low computational cost while still capturing the essential properties of TSCs~\cite{ap435.168431, prb95.245137, prb105.014507}. 

The Hamiltonian is given by
\begin{eqnarray}
H_{\rm TSC}=c_{x,y}^\dagger \mathcal{H}_0 c_{x,y} +\Delta(x,y) c_{x,y}^\dagger c_{x,y}^\dagger + \mathrm{H.c.},
\end{eqnarray}
where $c_{x,y}/c_{x,y}^\dagger$ represent the annihilation/creation operators for particles at position $(x, y)$, $c_{x,y}=(c_{x,y,\uparrow},c_{x,y,\downarrow})^T$, and $\Delta(x, y)$ denotes the superconducting pair potential. Here, $\mathcal{H}_0$ corresponds to the Hamiltonian governing surface states in a 3D topological insulator:
\begin{eqnarray}
\mathcal{H}_0=\hbar v_F (\boldsymbol{\sigma}\times\mathbf{k})\cdot \mathbf{e}_z+\frac{W\lambda}{2}\sigma_z|\mathbf{k}|^2,
\end{eqnarray}
with $v_F$ representing Fermi velocity. We introduce a mass term $W$ to address fermion doubling issues and restore a finite gap in Brillouin zone boundaries~\cite{prb95.245137}. In these equations, $\boldsymbol{\sigma}=(\sigma_x,\sigma_y,\sigma_z)$ refers to Pauli matrices acting on spin space and $\mathbf{k}=(k_x,k_y)$  represents momenta on the surface.

Expressed in the Nambu basis $C_{x,y}=(c_{x,y},c_{x,y}^\dagger)^T$, the TSC Hamiltonian can be written compactly as 
\begin{eqnarray}
  H_{\rm TSC}=C_{x,y}^\dagger \mathcal{H}_{\rm TSC} C_{x,y},
\end{eqnarray}
where the BdG Hamiltonian $\mathcal{H}_{\rm TSC}$ is given by
\begin{eqnarray}
\mathcal{H}_{\rm TSC}=h_1 \nu_z + h_2 +\Delta_R \nu_x - \Delta_I \nu_y,
\end{eqnarray}
with $h_1 = i \hbar v_F (\sigma_y \partial_x-\sigma_x \partial_y)$, and $h_2 = -\frac{W\lambda}{2}\sigma_z (\partial_x^2+\partial_y^2)$. Here, $\Delta$ represents a complex pairing potential with real and imaginary parts denoted as $\Delta_R$ and $\Delta_I$, respectively. The Pauli matrices $\nu_{x,y,z}$ act on the particle-hole space.

The dimensionless form of the BdG Hamiltonian is required to coherently integrate into the dimensionless TDGL equations, which can be deduced using the same procedure as in the previous section. The resulting final dimensionless Hamiltonian is given by:
\begin{eqnarray}\label{eqBdGHamiltonian}
\tilde{\mathcal{H}}_{\rm TSC} & = & i \tilde{v}_F (\sigma_y \tilde{\partial}_x - \sigma_x \tilde{\partial}_y) \nu_z - \tilde{W}\sigma_z (\tilde{\partial}_x^2\nonumber\\
& & + \tilde{\partial}_y^2) +\tilde{\Delta}_R\nu_x - \tilde{\Delta}_I\nu_y,
\end{eqnarray}
where the dimensionless quantities are defined as follows: 
\begin{eqnarray} 
\begin{aligned}
  \tilde{\mathcal{H}}_{\rm TSC} &=\frac{H_{\rm TSC}}{\Delta_0},             &
  \tilde{v}_F &=\frac{\hbar v_F}{\lambda\Delta_0},            \\
  \tilde{W} &=\frac{W}{2\lambda\Delta_0},             &
  \tilde{\Delta}_{R,I} &=\frac{{\Delta}_{R,I}}{\Delta_0}.                &
\end{aligned}
\end{eqnarray}
Here, $\Delta_0$ represents the amplitude of superconducting pair potential without vortex.

With the dimensionless $\tilde{\mathcal{H}}_{\rm TSC}$, the TDBdG Schr{\"o}dinger equation can be also written in a dimensionless form
\begin{eqnarray}
\label{eqTDBdG}
\tilde{\mathcal{H}}_{\rm TSC}(\tau)\phi_{\rm TSC}(\tau)=id\partial_\tau \phi_{\rm TSC}(\tau),
\end{eqnarray}

where $d=\frac{\hbar D}{\xi^2 \Delta_0}=\frac{\hbar}{\Delta_0 t_0}$ is a dimensionless constant that can be absorbed by rescaling the time variable to $\tau'=\tau/d$. After this rescaling, the equation becomes
\begin{eqnarray}
\tilde{\mathcal{H}}_{\rm TSC}(\tau')\phi_{\rm TSC}(\tau')=i\partial_{\tau'} \phi_{\rm TSC}(\tau').
\end{eqnarray}
This formulation represents the TDBdG formula used for numerical calculations of MZM braiding dynamics.

\begin{figure}
\centering
\includegraphics[width=8 cm]{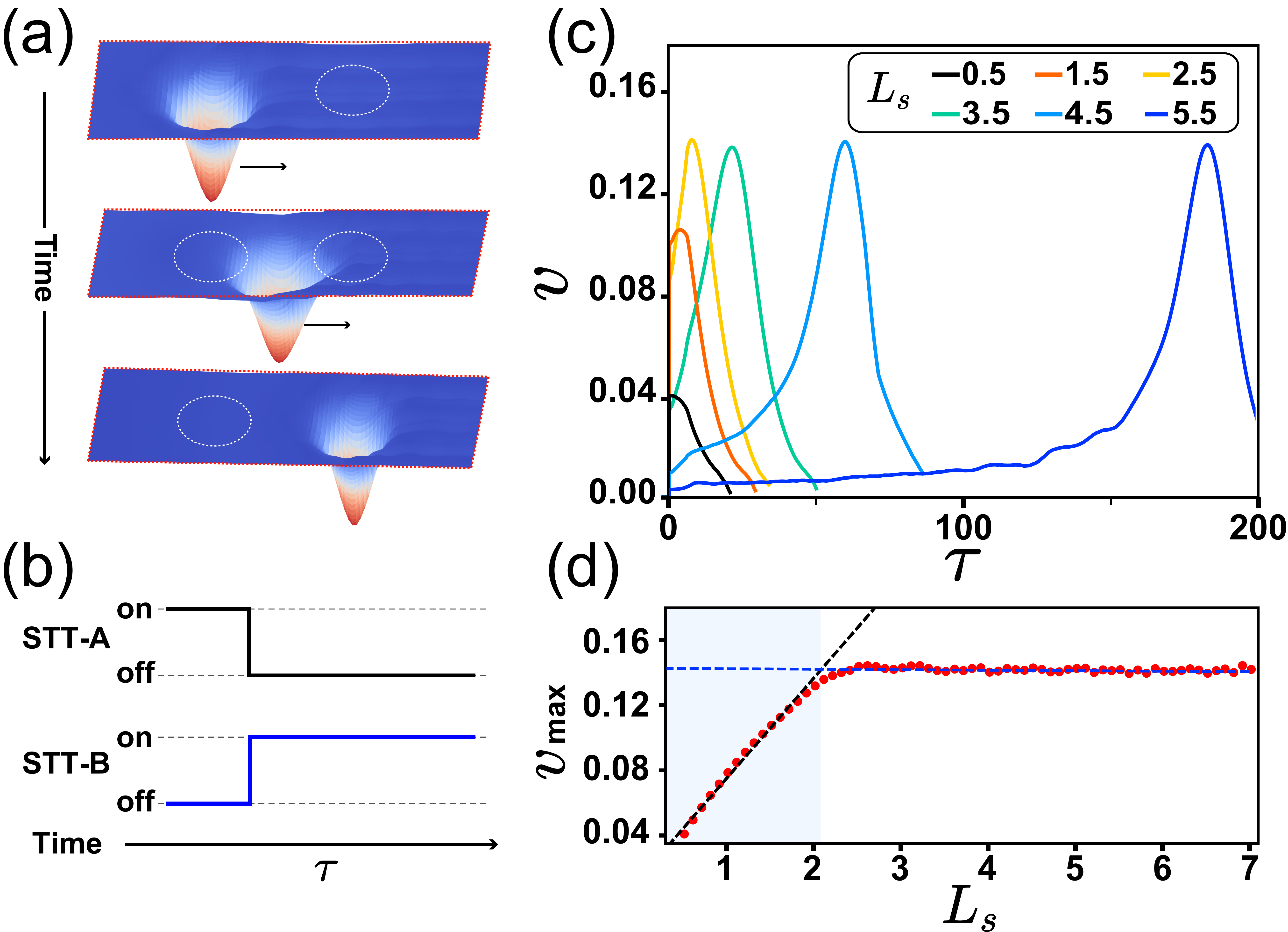}
\caption{\label{figDynamics} 
(Color online) Vortex dynamics induced by STTs. 
(a) Sketch of a single vortex's motion from the position of STT-A to STT-B shown in Fig. \ref{figDevice}(a). The vortex is represented by the order parameter.
(b) The operation sequences of the two STTs. 
(c) The velocity of the vortex $v$ versus the time $\tau$ at different STTs' distance $L_s$. 
(d) The maximum velocity $v_{max}$ versus $L_s$. The value of $v_{max}$ saturates when $L_s \gtrsim R_v$. The parameters used in the calculations are $\kappa=4$, $b_0=1.2$, $R_v=1.7$, $\sigma=1$, and $\xi=0.25$. Note that spatial parameters and temporal parameters are rescaled by $\lambda$ and $\xi^2/D$, respectively. Other parameters are the same values as in Fig.~\ref{figDevice}.
} 
\end{figure}

To incorporate vortices into the TDBdG Hamiltonian, it is necessary to modify the superconducting pair potential near a vortex pinned by the $i,j$-\textit{th} STT to
\begin{eqnarray}
\Delta_{i,j}(\boldsymbol{r},\tau)=\Delta_0 \tanh\frac{|\boldsymbol{r}-\boldsymbol{r_{i,j}}(\tau)|}{\xi}\mathrm{exp}(-i\mathrm{Arg}(\boldsymbol{r}-\boldsymbol{r_{i,j}}(\tau)))
\end{eqnarray}
where $\boldsymbol{r_{i,j}}(\tau)=(x_{i,j}(\tau),y_{i,j}(\tau))$ represents the central coordinates of the vortex pinned by the $i,j$-\textit{th} STT. Renormalize $\Delta_0$ to unity, the overall superconducting pair potential $\Delta(\boldsymbol{r},\tau)$ is given by
\begin{eqnarray}
\label{eqDelta}
\Delta(\boldsymbol{r},\tau)=\prod_{i,j} \Delta_{i,j}(\boldsymbol{r},\tau),
\end{eqnarray}
where the product is taken over all vortices.

\subsection{Simulation Procedure}
\label{subsecSimulation}

After introducing the basic components, a comprehensive simulation of the MZM braiding dynamics can be conducted. 

The initial step involves designing the braiding paths and determining the ON/OFF operation time sequences of the STTs based on specific braiding process. Subsequently, utilizing the defined braiding paths and STT time sequences, the TDGL equations yield the Cooper pair density $\rho(\tau)$, thereby enabling analysis of vortex dynamics. The trajectories of vortices $\{\mathbf{r}_{i,j}(\tau) = (x_{i,j}(\tau), y_{i,j}(\tau)) \mid i,j=1,2,3,\ldots \}$ are then extracted from $\rho(\tau)$ for all $\tau$ and passed to TDBdG equations through Eq.~(\ref{eqDelta}). By solving static BdG equations, we identify states of MZMs. Finally, by selecting an initial MZM state, we can obtain corresponding braiding dynamics along the designated paths.

\section{MZMs braiding}\label{secBraiding}

\subsection{Single Vortex Dynamics}
\label{subsecSingle}

To demonstrate the vortex-manipulation capability of STTs, we first simulate the motion of a single vortex under the dragging of STTs. 

As depicted in Fig.~\ref{figDynamics}(a), a vortex initially immobilized by device STT-A (indicated by the left dashed circle) undergoes drift towards the position of STT-B (right dashed circle) upon altering the ON/OFF states of both STT-A and STT-B, as shown in Fig.~\ref{figDynamics}(b). Eventually, the vortex reaches and settles into a position near STT-B, representing another stable state. We refer to this process as an elementary move. The TDGL simulation also indicates that the vortex dissipation time (after the corresponding STT being turned off) is much longer than the braiding time scale.

We investigated the dynamics of vortex elementary motion with varying distances $L_s$ between the two STTs. The displacements of the vortex, denoted as $D(\tau)$, were first extracted from the simulation results obtained by solving the TDGL equations. Subsequently, we determined the velocities of the vortex, represented as $v(\tau)(=dD(\tau)/d\tau)$. As illustrated in Fig.~\ref{figDynamics}(c), for small $L_s$, immediate capture of the vortex by STT-B occurred after its release by STT-A due to its proximity within the effective region of stray field generated by STT-B, resulting in a rapid acceleration in velocity. Conversely, for large $L_s$, initially outside reach of STT-B's stray field, a transient wandering phase was observed before eventual capture by STT-B.

Notably, the maximum velocity ($v_{max}$) reaches saturation when $L_s$ exceeds a certain threshold value. This behavior is further elucidated in Fig.~\ref{figDynamics}(d), where data points corresponding to different values of $L_s$ are collected to demonstrate two distinct patterns with fitted lines. The maximum velocity of vortex motion is intricately linked to both the speed of braiding operations and adiabaticity. Besides, we will also see in Subsec.~\ref{subsecBraiding} that large accelerations of vortex (small $L_s$) leads to a high braiding error.

As mentioned earlier, for a type-II superconductor, the GL parameter $\kappa$ should satisfy $\kappa>1/\sqrt{2}$. In Fig.~\ref{figDynamics}, we set the value of $\kappa$ to 4. It is worth noting that our model remains valid across a wide range of $\kappa$, encompassing both the clean superconductor regime (small values of $\kappa$) and the iron-based superconductor regime (large values of $\kappa$).

\begin{figure}
\centering 
\includegraphics[width=8 cm]{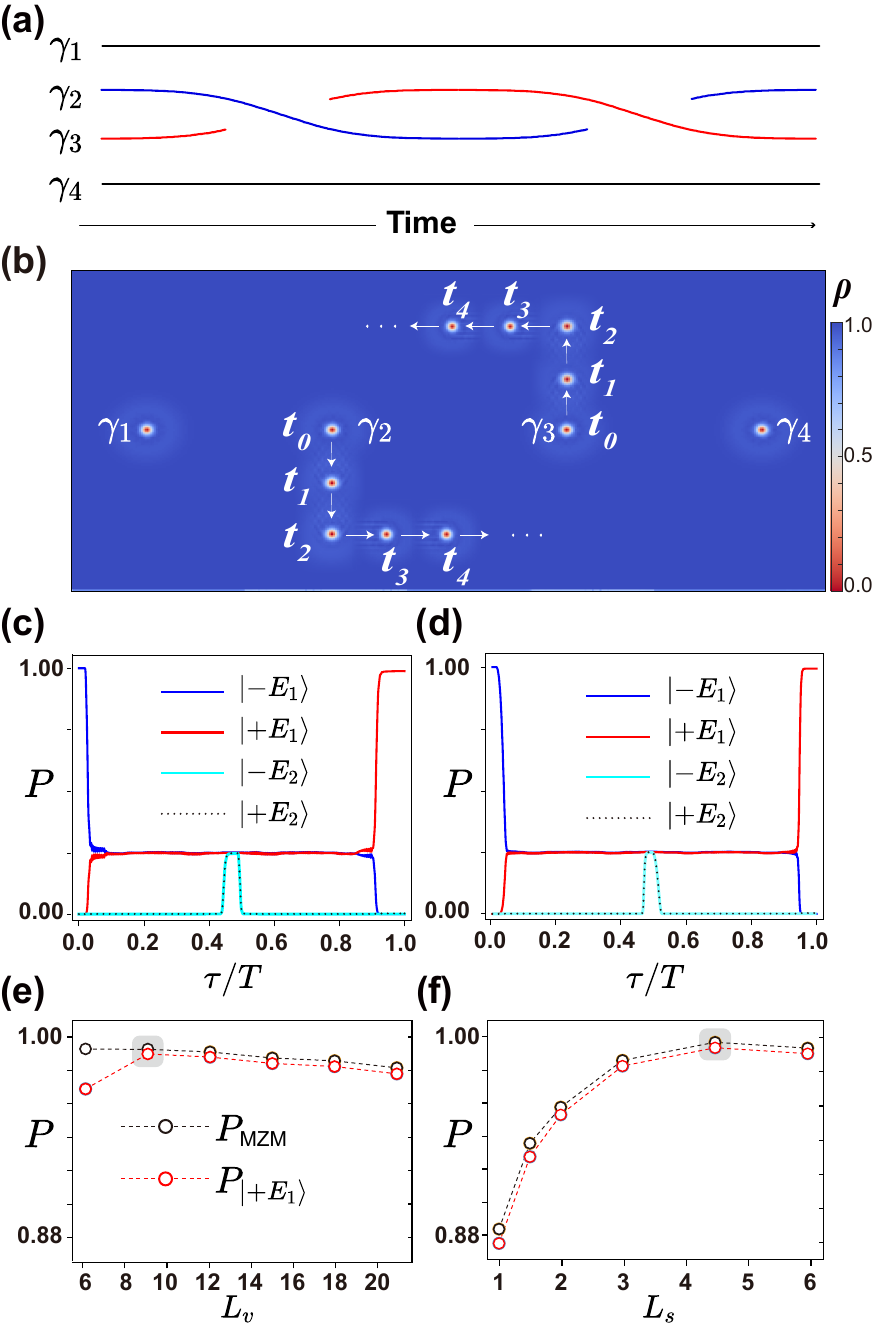}
\caption{\label{figBraiding} 
(Color online) Braiding dynamics of MZMs facilitated by the STT-matrix. 
(a) Worldlines illustrating the braiding operator $\mathbb{B}_X=b_2b_2$ for four MZMs $\gamma_{i=1,2,3,4}$. 
(b) Cooper pair density distribution $\rho(x,y)$ in TSC layer with depicted vortices, where the two middle vortices are driven along a programmable path indicated by arrows using STTs. Vortices associated with $\gamma_{2,3}$ are overlaid on the background following the braiding trajectory.
(c) Evolution of $P_{|\pm E_i\rangle}, i=1,2$ (defined in the text) at $L_v=9$ and $L_s=3$. 
(d) Evolution of $P_{|\pm E_i\rangle}, i=1,2$ at $L_v=18$ and $L_s=4.5$. 
(e) Variation of both $P_{\rm MZM}$ and $P_{|+E_1\rangle}$ as functions of $L_v$, while keeping $L_s$ fixed at 3. 
(f) Dependence of both $P_{\rm MZM}$ and $P_{|+E_1\rangle}$ on varying values of $L_s$, with fixed value of $L_v = 18$.
The state $|-E_1\rangle$ is chosen as the initial state in all the calculations. Other parameters are the same values as in Fig. \ref{figDevice}.
} 
\end{figure}

\subsection{Braiding}

\label{subsecBraiding}

The previous subsection demonstrated the STT's ability to drive the vortex, and it is natural to extend this capability in an STT-matrix. By properly arranging the status of the STTs, it becomes possible to freely move the vortex along the trajectory defined by of the STT lattice. The collection of these statuses is referred to as a time sequence, which consists of status functions for all STTs $\{s_i(\tau)\}$ defined in section~\ref{secModel}. This implies that any braiding operation can be implemented by utilizing the interaction between STTs and vortices. To illustrate this point, it would be beneficial to begin with an introduction of non-Abelian anyon braiding.

In the TQC proposal, the computational process of manipulating non-Abelian anyons is represented by their space-time trajectories, known as worldlines. These worldlines can be formally described using the braiding group framework. According to this theory, a general braiding operation $\mathbb{B}$ can be decomposed into a combination of elementary operations $\mathbb{B}=b_ib_j \cdots b_k$, where $i,j,\cdots k\in\pm\{1,\ldots,n-1\}$ (with $n$ being the number of non-Abelian anyons). The elementary operation $b_i$ ($i>0$), referred to as a generator of the braiding group, represents a counterclockwise exchange between the $i$-th and $(i+1)$-th non-Abelian anyons, while the corresponding operator $b_{-i}$ is the inverse of $b_i$ representing a clockwise braiding. This decomposition ensures that any general braiding operation can be achieved by performing all necessary generators, which can be facilitated by digitally controlling the STT-matrix.

In practice, the braiding operation requires two essential preparatory steps. Firstly, it is necessary to establish a mapping between the braiding worldlines and the corresponding braiding paths in the STT lattice. Secondly, one must design appropriate time sequences for the STTs along these braiding paths, as illustrated in Figs.~\ref{figBraiding}(a)-(b).

To demonstrate the braiding of MZMs, we calculate the dynamics of MZM braiding using the TDBdG equations. The system under consideration consists of four vortices with even interval, with each vortex hosting an individual MZM denoted as $\gamma_{i=1,2,3,4}$. For illustrative purposes, we choose the braiding operation $\mathbb{B}_X=b_2b_2$ (i.e., exchange the positions of $\gamma_2$ and $\gamma_3$ twice, which is an effective $X$-gate), which is selected for its simplicity and ability to yield a distinct signature in the braiding dynamics.

In the language of TQC, four MZMs can be encoded into one logical qubit defined in terms of fermionic modes, specifically the even subspace. For example, we have 
\begin{eqnarray}
\{|\bar{0}\rangle=|00\rangle, |\bar{1}\rangle=|11\rangle=c_1^\dagger c_2^\dagger |00\rangle\},
\end{eqnarray}
where $c_1=\frac{1}{2}(\gamma_1+i\gamma_2)$, $c_2=\frac{1}{2}(\gamma_3+i\gamma_4)$ represent the fermionic operators. The parity of a pair of MZMs can be defined as the fermion number associated to the fermionic operator. Here, $|\bar{0}\rangle$ and $|\bar{1}\rangle$ denote the logical qubit. By performing a braiding operation $\mathbb{B}_X$, MZMs $\gamma_2$ and $\gamma_3$ acquire a $\pi$-phase due to their non-Abelian statistics~\cite{prb91.174305,prb103.054504,prb105.014507}, resulting in $\gamma_2\to-\gamma_2$ and $\gamma_3\to-\gamma_3$, while leaving $\gamma_{1}$ and $\gamma_{4}$ unchanged. Consequently, the associated logical qubit undergo a transformation from $|\bar{0}\rangle \to |\bar{1}\rangle$ and $|\bar{1}\rangle \to |\bar{0}\rangle$, which essentially corresponds to a quantum NOT gate operation on the logical qubit representation. This becomes evident by noting that after braiding, the fermionic operators transform as follows: 
\begin{eqnarray}
c_1=\frac{\gamma_1+i\gamma_2}{2}\to \frac{\gamma_1-i\gamma_2}{2}=c_1^\dagger,  \\
c_2=\frac{\gamma_3+i\gamma_4}{2}\to \frac{-\gamma_3+i\gamma_4}{2}=-c_2^\dagger.
\end{eqnarray}

The nontrivial transformation of the operators can also be elucidated in the picture of single-particle spectrum within the MZM subspace. The states in the MZM subspace can be obtained by solving the static BdG Hamiltonian Eq. (\ref{eqBdGHamiltonian}). Theoretically the MZMs are the states appear inside the superconductor gap with zero energy.
However, due to the overlaps of their wavefunctions, the energies split to values close to zero. The particle-hole symmetry ensures the simultaneous existence of $|\pm E\rangle$ states. The splitting of energy depends on the distance between the MZMs. Therefore, by singling out the two pairs of particle-hole symmetric subgap eigenstates, we have identified and collected a set of states constituting the MZM subspace, denoted as $\{|\pm E_{i}\rangle (i=1,2)\}$. 

After the braiding operation $\mathbb{B}_{X}$, the states undergoes a transition: ${ |\pm E_i \rangle} \rightarrow {| \mp E_i \rangle}, (i = 1 , 2)$. This nontrivial transition can be understood as follows. Suppose we start from the many-body state $|\bar{0}\rangle=|00\rangle$, which annihilated by the operators $c_1$ and $c_2$ ($c_1|\bar{0}\rangle=c_2|\bar{0}\rangle=0$). This corresponds to the occupied single-particle states being $|-E_1\rangle$ and $|-E_2\rangle$, as they each satisfy the annihilation conditions $c_1|-E_1\rangle=0$ and $c_2|-E_2\rangle=0$. Take $|-E_1\rangle$ as an example. It should satisfy the constraints before and after the braiding, in order to deduce the final state. Concretely, the initial state is annihilated by $c_1$ ($c_1|-E_1\rangle=0$), but after the braiding, the operator becomes $c_1\to c_1^\dagger$, thus requiring the final state be annihilated by $c_1^\dagger$ ($c_1^\dagger|f\rangle=0$), which means the final state is $|f\rangle=|+E_1\rangle$. Collecting each case of these states, it is implied that the braiding operation $\mathbb{B}_X$ causes a state transition as $|\pm E_i\rangle\to|\mp E_i\rangle,i=1,2$.

To characterize the fidelity of braiding operation, we focus on two physical quantities. The first physical quantity is the state transition probability, defined as 
\begin{eqnarray}
P_{\Phi}(\tau)=|\langle\phi(\tau)|\Phi(0)\rangle|^2,
\end{eqnarray}
where $|\phi(\tau)\rangle$ represents the braided state and $|\Phi(0)\rangle$ denotes the initial MZM state labeled as $|\pm E_i\rangle, i=1,2$ at time $\tau=0$. This probability quantifies the likelihood of a braided state projecting onto the encoded space constituted by MZMs. The second physical quantaty is the leakage out of the MZM subspace during operation, which can be denoted as $1-P_{\rm MZM}$, where $P_{\rm MZM}=\sum_{\Phi} P_{\Phi}(\tau), \Phi \in \{|\pm E_i \rangle | i=1,2\}$.

By applying the TDBdG equations Eq. (\ref{eqTDBdG}) and choosing the initial state as $\phi(\tau=0)=|-E_1\rangle$, the entire braiding process of the braided state $\phi(\tau)$ can be calculated. The state transition probabilities throughout the entire $\mathbb{B}_X$ braiding process are depicted in Figs.~\ref{figBraiding}(c) and (d). It is evident that the process initiates with a high $P_{|-E_1\rangle}$ and ends with a high $P_{|+E_1\rangle}$. Ideally, $P_{|+E_1\rangle}$ should reach unity after $\mathbb{B}_X$, while $P_{\rm MZM}$ should remain at unity for the entirety of the operation. 

However, the braiding fidelity can be influenced by two factors. Firstly, a finite coupling strength between MZMs, denoted as $E_M$, can transform MZMs into fermionic modes and cause loss of their non-Abelian statistics. The finite value for $E_M$ primarily results in a reduction in $P_{|+E_1\rangle}$. Secondly, the nonadiabatic effects arising from vortex dynamics significantly impact the system. The velocity nonuniformity induced by vortex motion can be regarded as a perturbation to MZMs, particularly during the processes of vortex releasing and repinning. When $L_s$ is small, the rapid changes in vortex dynamics [as illustrated in Fig.~\ref{figDynamics}(c)] lead to pronounced fluctuations in velocity. This results in substantial leakage from the MZM computational subspace into the quasi-continuum, thereby reducing both $P_{\rm MZM}$ and $P_{|+E_1\rangle}$.

To investigate the impact of finite $E_M$ and nonadiabaticity on the braiding process in the STT-matrix scheme, we examined the dependence of $P_{|+E_1\rangle}$ and $P_{\rm MZM}$ on $L_v$ and $L_s$. Here, $L_v$ represents the interval between $\gamma_i$ and $\gamma_{i+1}$, serving as an indicator for $E_M$. As depicted in Fig.~\ref{figBraiding}(e), when $L_v$ is small, there is a significant reduction in $P_{|+E_1\rangle}$ while maintaining a high value for $P_{\rm MZM}$. Conversely, when $L_v$ becomes large enough to neglect the effect of $E_M$, both $P_{|+E_1\rangle}$ and $P_{\rm MZM}$ approach unity. With further increase in $L_v$, gradual drops are observed in both $P_{|+E_1\rangle}$ and $P_{\rm MZM}$ due to an increase in elementary moves. The value of $L_s=3$, remains constant for all data points presented in Fig.~\ref{figBraiding}(e). In Fig.~\ref{figBraiding}(f), we fix $L_v=18$, while varying $L_s$. It can be clearly seen that both $P_{|+E_1\rangle }$ and $P_{\text{MZM}}$ increase with increasing $L_s$. This behavior arises from the fact that smaller $L_s$ will result in larger accelerations experienced by vortices and cause larger leakage (recalling the analysis in Sec.~\ref {subsecSingle}). Additionally, smaller values of $L_s$, require more elementary moves.


\subsection{Fusion}

\label{subsecFusion}
The fusion of anyons~\cite{ap303.2,rmp80.1083,bookIntroToTQC} is a crucial step in bringing them together to decode information in the computational basis, which plays an integral role in the read-out process of TQC. Moreover, the utilization of fusion alone (without braiding) can also facilitate measurement-based TQC~\cite{prl101.010501}.

\begin{figure}
  \centering 
  \includegraphics[width=8 cm]{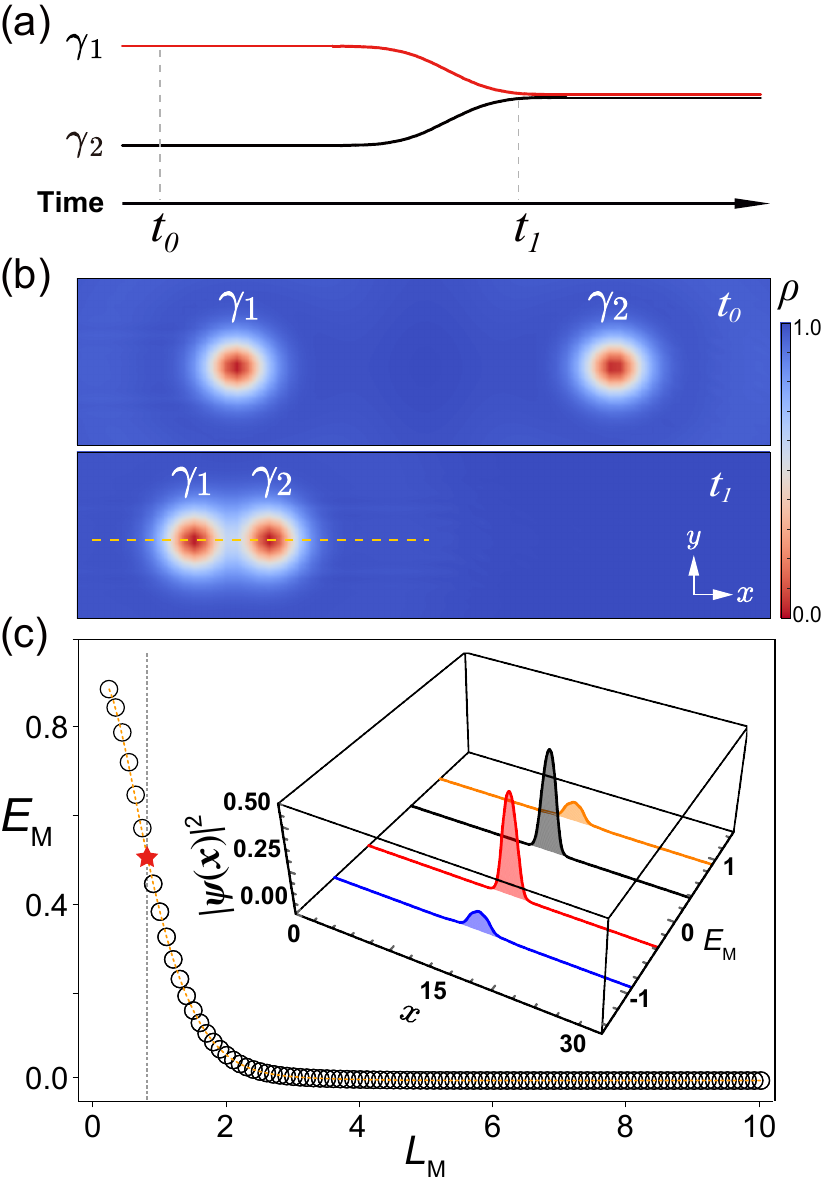}
  \caption{\label{figFusion} 
(Color online) Fusion of two MZMs within vortices induced STTs. 
(a) Worldlines depicting the fusion process of two MZMs are presented. 
(b) Simulated Cooper pair density $\rho(x,y)$ illustrates the dynamics of vortex fusion. 
(c) Energy splitting $E_M$ between two MZMs against their separation distance $L_M$. The inset displays the probability density distribution of MZMs at $L_M \approx \lambda$.
  } 
\end{figure}

In the STT-matrix braiding scheme, the fusion operation can be easily performed by bringing together the vortices hosting MZMs that are intended to be fused, as illustrated in Figs.~\ref{figFusion}(a) and (b). The controlled movement of these two MZMs can be achieved through the application of STTs (see Sec.~\ref{subsecSingle}), ultimately leading them to approach each other. Subsequently, the MZMs are pinned at a pair of neighboring STTs set in an ON state. 

Remarkably, we observed that even if one of the two STTs is switched OFF, its associated vortex will still migrate towards the remaining ON-STT. In essence, this implies that two vortices can symmetrically localize around a single ON-STT. However, due to their repulsive interaction~\cite{bookIntroToSC}, these vortices remain separated by a distance of around $\lambda$.

After the STT operations, the fusion of the two MZMs within the neighboring vortices occurs, resulting in an observable finite energy splitting denoted as $E_M$. The dependence of $E_M$ on the distance between MZMs, referred to as $L_M$, is illustrated in Fig.~\ref{figFusion}(c). Due to the relatively large Fermi velocity in metal, the MZM wavefunction does not show oscillations along with $L_M$~\cite{DasSarma2012}.

The quantum information stored in the fused MZMs can be extracted through charge flipping or other related physical effects, as proposed in previous works on read-out methods for 2D-TSC systems~\cite{prapp12.054035, prb105.014507}. These methods are compatible with the STT-matrix scheme; however, their implementation is beyond the scope of this study.

\section{Conclusion}
\label{secConclusion}

In conclusion, we investigate the braiding of MZMs in 2D TSC vortices using an STT-matrix platform. To validate our proposal, we develop a theoretical simulation method based on the TDGL theory and TDBdG equations. By incorporating the STT-matrix into the TDGL equations through modeling local magnetic stray fields, we simulate individual MZM manipulation within vortices driven by STTs. Subsequently, this method is employed to study vortex dynamics induced by STTs and further explore MZM braiding dynamics using TDBdG equations. We demonstrate that manipulating the STT-matrix enables achieving MZM movement, braiding, and fusion operations. Additionally, we show that nonadiabaticity and finite coupling of MZMs significantly impact their braiding dynamics, which can be adjusted in our proposed scheme. Our approach is general and independent of concrete TSC models, such as the Fu-Kane model, semiconductors with Rashba spin-orbit interaction, and iron-based superconductor model. We believe the studies on the MZMs braiding in the vortex states of TSC will advance TQC.

\vspace{5pt}

\textbf{Acknowledgment.} This work was supported by the Innovation Program for Quantum Science and Technology (Grant No. 2021ZD0302401), the Hunan Provincial Science Foundation for Distinguished Young Scholars (Grant No.2021JJ10043), and the Open Research Fund from State Key Laboratory of High Performance Computing of China (HPCL) (Grant No.201901-09 and No.2023-KJWHPCL-07.

\vspace{5pt}
G.Y.H and J.B.F contribute equally to this work.


\appendix 

\renewcommand{\thefigure}{A\arabic{figure}}
\setcounter{figure}{0} 

\section{Derivation of TDGL equations}
\label{secAppendixA}

The derivation of dimensionless TDGL equations is presented in detail in this section. We commence with the dimensional TDGL equations~\cite{aam115.63,bookShortcut}.
\begin{eqnarray}
  \label{eqTDGLDimension2}
\frac{\hbar^2}{2mD}(\frac{\partial}{\partial t}+i\frac{q}{\hbar}\Phi)\Psi
& = & -\frac{1}{2m}(\frac{\hbar}{i}\nabla-q\mathbf{A})^2\Psi+\alpha\Psi\nonumber\\
&   & -\beta|\Psi|^2\Psi,\\
\sigma(\frac{\partial \mathbf{A}}{\partial t}+\nabla\Phi)
& = & \frac{q\hbar}{2mi}(\Psi^*\nabla\Psi-\Psi\nabla\Psi^*)\\
&   & -\frac{q^2}{m}|\Psi|^2 \mathbf{A}-\frac{1}{\mu_0}\nabla\times(\nabla\times \mathbf{A}-\mathbf{B}).\nonumber
\end{eqnarray}
Parameter definitions can be found in the main texts.

We define:
\begin{eqnarray} \begin{aligned}
  \partial_t &=\frac{D}{\xi^2}\partial_\tau,                  &
  \nabla &=\frac{\tilde{\nabla}}{\lambda},                    &
  \Phi &=\frac{\hbar D \kappa}{q\xi^2} \phi,                  &
  \Psi &=\sqrt{\frac{\alpha}{\beta}}\psi,\\                   
  \sigma &=\frac{\tilde{\sigma}}{\mu_0 D \kappa^2},           &
  \mathbf{A} &=\frac{\hbar}{q\xi}\tilde{\mathbf{A}},          &
  \mathbf{B} &=\frac{\hbar}{q \kappa \xi^2} \mathbf{b}.   & 
\end{aligned} \end{eqnarray}
Substituting the parameters above into Eq.~\ref{eqTDGLDimension} (or Eq.~\ref{eqTDGLDimension2}), we get a dimensionless form:
\begin{eqnarray}
(\frac{\partial}{\partial \tau}+i\kappa\phi)\psi & = & -(\frac{i}{\kappa}\tilde{\nabla}+\tilde{\mathbf{A}})^2\psi+\psi-|\psi|^2\psi,\\
\sigma(\frac{\partial \tilde{\mathbf{A}}}{\partial \tau}+\tilde{\nabla}\phi) & = & \frac{1}{2i\kappa}(\psi^*\tilde{\nabla}\psi-\psi\tilde{\nabla}\psi^*)-|\psi|^2 \tilde{\mathbf{A}}\nonumber\\
& & -\tilde{\nabla}\times(\tilde{\nabla}\times \tilde{\mathbf{A}}-\mathbf{b})
\end{eqnarray}

To further simplify the TDGL equations, we perform the gauge transformation on the electromagnetic potential.
\begin{eqnarray}
\begin{array}{rcl}
\psi' &=& \psi e^{i\kappa \chi},
\\
\tilde{\mathbf{A}}' &=& \tilde{\mathbf{A}}+\tilde{\nabla} \chi,
\\
\phi' &=& \phi-\partial_\tau \chi,
\end{array}
\end{eqnarray}
where $\chi=\chi(x,y,z,t)$ is an arbitrary real function. Fixing the gauge by choosing $\phi=\partial_\tau \chi$, the electric potential $\phi'$ will vanish globally. 

Since TDGL equations are gauge invariant, they now have the form:
\begin{eqnarray}
\partial_\tau \psi & = & -(\frac{i}{\kappa}\tilde{\nabla}+\tilde{\mathbf{A}})^2\psi+\psi-|\psi|^2\psi,\\
\sigma \partial_\tau \tilde{\mathbf{A}} & = & \frac{1}{2i\kappa}(\psi^*\tilde{\nabla}\psi-\psi\tilde{\nabla}\psi^*)-|\psi|^2 \tilde{\mathbf{A}}\nonumber\\
& & -\tilde{\nabla}\times(\tilde{\nabla}\times \tilde{\mathbf{A}}-\mathbf{b}).
\end{eqnarray}
Redefine 
\begin{eqnarray}
\begin{array}{rcl}
\tilde{\nabla} &\to& \nabla,
\\
\tilde{\mathbf{A}} &\to& \mathbf{A},
\\
\tilde{\sigma} &\to& \sigma,
\end{array}
\end{eqnarray}
and get the final form of TDGL equations in the main text: 
  \begin{eqnarray}
  \partial_\tau \psi & = & -(\frac{i}{\kappa}\nabla+\mathbf{A})^2\psi+\psi-|\psi|^2\psi\\
 \sigma \partial_\tau \mathbf{A} & = & \frac{1}{2i\kappa}(\psi^*\nabla\psi-\psi\nabla\psi^*)-|\psi|^2 \mathbf{A}\nonumber\\
  & & -\nabla\times(\nabla\times \mathbf{A}-\mathbf{b})
 \end{eqnarray}
\vspace{10pt}


\section{Experimental Feasibility}
\label{secAppendixFeasibility}

As commercially mature spintronic components, STT devices provide a robust foundation for the proposed MZM braiding scheme. STT technology has advanced rapidly since the first experimental demonstrations of STT-induced magnetization switching in metal multilayers \cite{katine2000} and magnetic tunneling junctions (MTJs) \cite{Zutic2004}. Its commercialization in non-volatile memories (MRAMs) \cite{apalkov2016} and scalability to gigabit integration on 22 nm/28 nm nodes \cite{2018Ong,2019Aggarwal} underscores the practical potential of an STT-matrix for MZM control.

Regarding key device parameters and material compatibility, the proposed scheme is compatible with experimentally validated 2D-TSCs. Integration of the STT-matrix with 2D-TSC materials can be achieved via flip-chip packaging technology ~\cite{FlipChip2016, Banda2008} or polymer-assisted 2D material transfer methods~\cite{Schranghamer2021}. Validated 2D-TSC materials include iron-based superconductors~\cite{Wang2018, prx8.041056, Kong2019, Zhu2019, Zhu2021, Kong2021nc, Li2022, Ge2023} and topological insulator/s-wave superconductor hybrids~\cite{prl114.017001, prl116.257003}. Such materials host vortex-bound MZMs and support the vortex dynamics modeled here. The diameter of STT-matrix elements can be scaled down to sub-10 nm \cite{watanabe2018}, matching the superconducting coherence length of iron-based TSCs~\cite{Wang2018, Hu2024}. This dimensional compatibility ensures localized stray fields can precisely pin individual vortices. It should be noted that current STT material fabrication processes are primarily optimized for fast memory devices; therefore, they require optimization to generate moderate magnetic fields suitable for MZM control.

For dynamic performance, the magnetization flip time of STT free layers has reached 200 ps$\sim$10 ns \cite{Thomas2019}. This guarantees that the single vortex manipulation time is significantly shorter than the theoretical coherence time of the topological qubits~\cite{Karzig2021}. Furthermore, thermal effects are negligible: STT switching energy is $\sim 10$ fJ per operation at room temperature \cite{chen2016,krizakova2022}, increasing only slightly at cryogenic temperatures while remaining manageable. Given that dilution refrigerators provide $\sim 10\ \mu$W of cooling power at mK temperatures, the STT-matrix can operate at gigahertz frequencies without disrupting superconductivity.

\begin{figure}
\centering
\includegraphics[width=6.5cm]{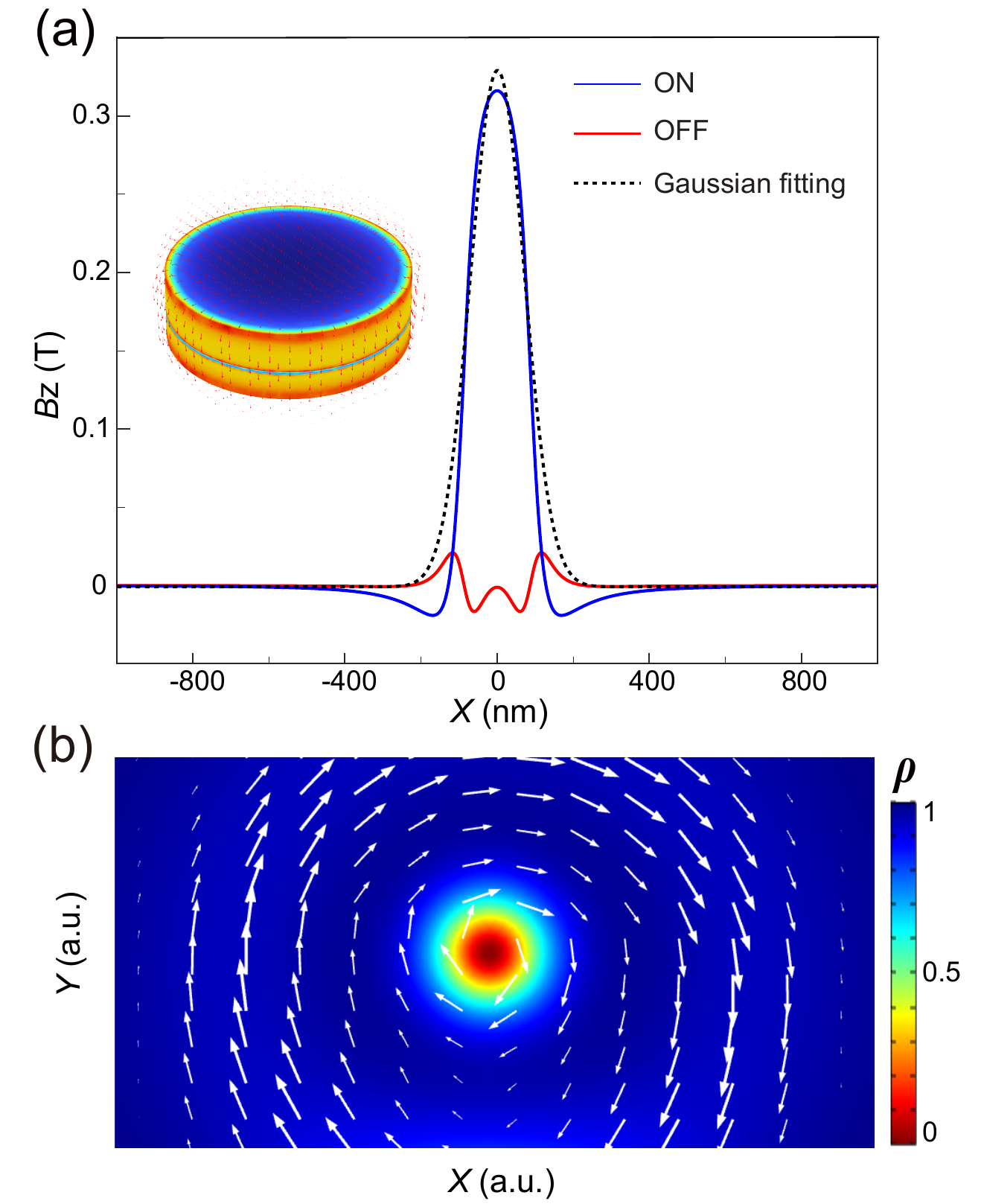}
\caption{(Color online) FEA simulation of an STT device. (a) Simulated stray field of an STT device in the ON and OFF states. The linecuts are the $B_z$-component of the stray field along a horizontal axis located 40 nm below the STT: the $B_z$ in the ON state (blue line), the Gaussian fitted $B_z$ in the ON state (dotted black line), and the $B_z$ in the OFF state (red line). The inset depicts a 3D simulation of the STT in the ON state. The magnetization intensities of the reference layer and free layer are set to $1.7\times10^6$ A/m~\cite{Zhou2019}. (b) Simulated Cooper pair density $\rho(x,y)$ by the FEA data [the blue line in (a)]. The current density vectors surrounding the vortex are indicated by white arrows.}
\label{AppFig1}
\end{figure}

\section{FEA Modeling of STT devices}
\label{secAppendixFEA}

To characterize the stray field of the STT device, we conducted a Finite Element Analysis (FEA) simulation for a single STT. 

In this FEA method, the STT device is modeled as a simplified cylindrical structure consisting of three layers: a 30-nm-thick reference layer on top, followed by a 2-nm-thick tunneling layer separating it from the bottom 20-nm-thick free layer. The radius of the cylindrical structure is 90 nm. It should be noted that real STT devices often have complex multiple-layer structures; however, our FEA model captures the essential local distribution of stray fields.

To simulate the distribution of stray field in the ON and OFF states, we manually manipulate the magnetization direction of the free layer to align it either parallel or antiparallel with the reference layer. The linecuts of the simulated $B_z$-component of the stray field along a horizontal axis located 40 nm below the STT is illustrated in Fig.~\ref{AppFig1}(a). It can be seen that for the ON state, there is a significant presence of stray field at the STT center, which can be approximately fitted by a Gaussian function in the TDGL simulation. Conversely, in the OFF state, magnetic field strength is considerably weaker and can be regarded as zero during the TDGL simulation.

\begin{figure}
\centering
\includegraphics[width=8 cm]{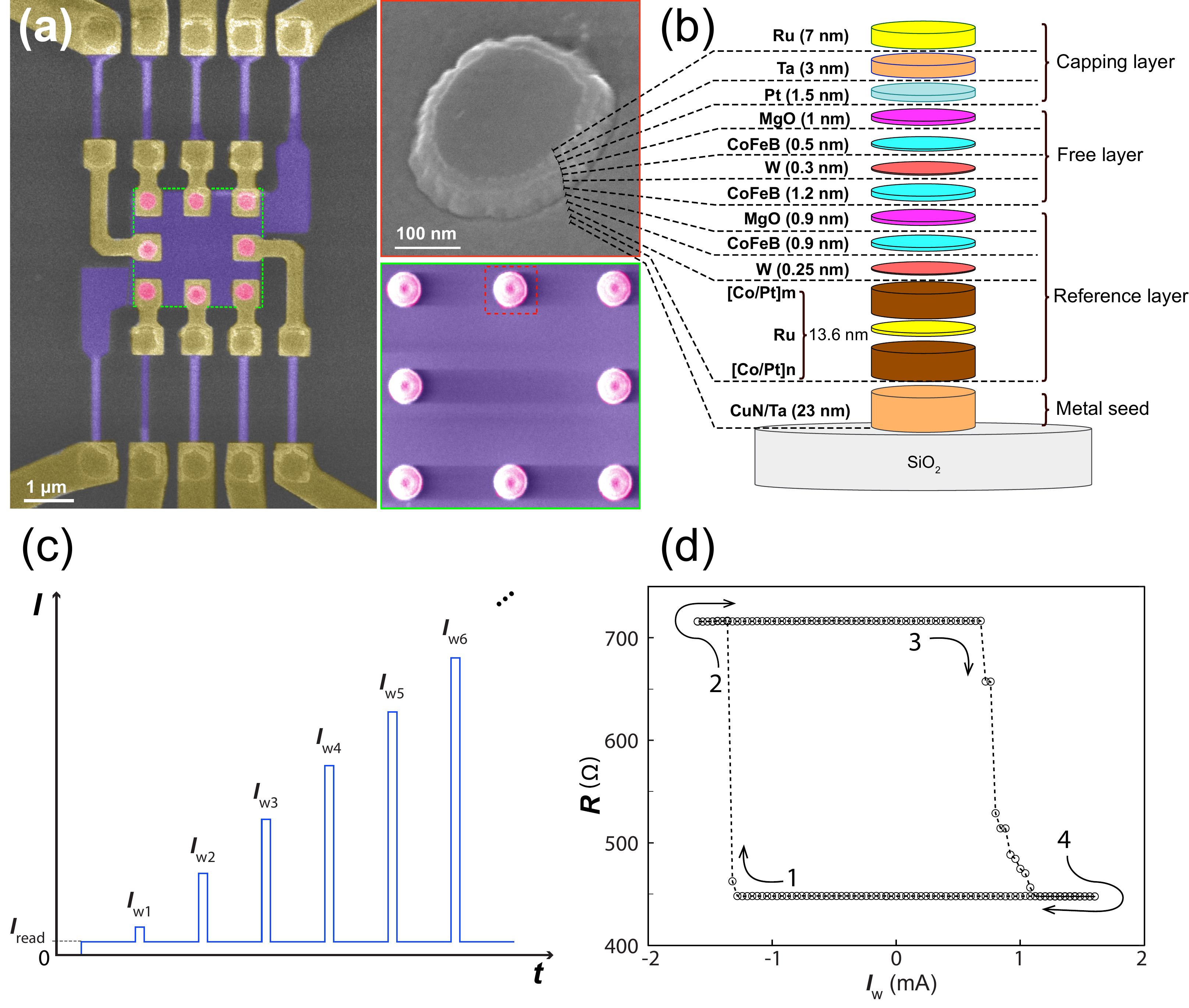}
\caption{(Color online) An experimental demonstration of STT-units. (a) A false-colored scanning electron microscope (SEM) image of an STT device. The pink-colored nanopillars in the middle are the STT unit, the pale-yellow parts and the purple parts are top and bottom electrodes, respectively. (b) Structure of the MTJ-material stack. (c) The write and read pulse current for characterizing the STT-unit. The amplitude of read current $I_{\rm read}$ remains a constant. The amplitude of write current $I_{wn}$ sweeps in both directions. (d) Device resistance versus write current pulse, sweeping in both directions. Numbers and arrows show the perpendicular magnetization transitions from AP to P states or the opposite situation.}
\label{AppFig2}
\end{figure}

From the linecuts, we observe a deviation in the negative part of $B_z$ from the Gaussian fitted lines adjacent to the peak. This deviation arises due to the closed nature of magnetic field lines in the STT device. However, this deviation does not cause any other consequences, such as the formation of a vortex-antivortex pair, since the reversed magnitude of $B_z$ is too small. To demonstrate this, we simulate a vortex by accounting the negative part of $B_z$ using FEA data in TDGL equations. The resulting vortex shown in Fig.~\ref{AppFig1}(b) exhibits no difference compared to the results obtained using Gaussian magnetic fields shown in the main text.

\section{An example of STT-matrix device}
\label{secAppendixDevice}

Here, we demonstrate the fabrication and measurements of a very-first-stage STT device. As the SEM image shown in Fig.~\ref{AppFig2}(a), an STT-Matrix with eight STT-units are fabricated. The STT-arrays are based on a multilayer MTJ material stack Fig.~\ref{AppFig2}(b). Top electrodes and a (global) bottom electrode are connected to the STT-array. In fact, these STT-devices are similar to those used in non-volatile memories. We characterized the STT-matrix using a write current pulse with various amplitude $I_{w1},I_{w2},\ldots$ and a following read current $I_{\rm read}$ of 10 $\mu$A in amplitude Fig.~\ref{AppFig2}(c). The write and read pulse durations are 1 ms and 300 ms, respectively. As shown in Fig.~\ref{AppFig2}(d), the resistance of the measured STT is bistable, and can be switched when the write pulse goes beyond a critical point. This is the so called STT-effect. The switching of the STT resistance indicates a switch of the free-layer magnetization. As just a demonstration, the measurements are performed at room temperature. Parameters like device compatibility at low temperature, the strength of stray-fields and the effect of write pulse Joule heat etc. need to be further investigated.

\bibliography{vortexBib}

\end{document}